\documentclass[aip,reprint]{revtex4-2}

\usepackage{algorithm}
\usepackage{algpseudocode}
\usepackage{lipsum}
\usepackage{amsmath}
\usepackage{amssymb}
\usepackage{graphicx}
\usepackage{gensymb}
\usepackage{tabularx}
\usepackage{float}
\floatstyle{plain}
\restylefloat{table}  
\usepackage{xcolor}
\usepackage{mathtools}
\usepackage{hyperref} 
\usepackage[capitalise]{cleveref}
\DeclareUnicodeCharacter{2009}{ }
\DeclareUnicodeCharacter{2192}{$\rightarrow$}
\DeclareUnicodeCharacter{2032}{'}
\DeclareUnicodeCharacter{03B1}{$\alpha$}
\DeclareUnicodeCharacter{03B3}{$\gamma$}
\restylefloat{table}
\graphicspath{ {images/} }

\begin{document}
	\title{Fusion alpha particle momentum deposition in thermonuclear burn dynamics}
	\author{A. J. Crilly}\email{ac116@ic.ac.uk}
	\affiliation{Centre for Inertial Fusion Studies, The Blackett Laboratory, Imperial College, London SW7 2AZ, United Kingdom}
    \affiliation{I-X Centre for AI In Science, Imperial College London, White City Campus, 84 Wood Lane, London W12 0BZ, United Kingdom}
	\author{B. D. Appelbe}
	\affiliation{Centre for Inertial Fusion Studies, The Blackett Laboratory, Imperial College, London SW7 2AZ, United Kingdom}
    \author{E. A. Ferdinandi}
	\affiliation{Centre for Inertial Fusion Studies, The Blackett Laboratory, Imperial College, London SW7 2AZ, United Kingdom}
    \author{S. T. O'Neill}
	\affiliation{York Plasma Institute, School of Physics, Engineering and Technology, University of York, Heslington, York YO10 5DD, United Kingdom}
    \author{H. Biragnet}
	\affiliation{Centre for Inertial Fusion Studies, The Blackett Laboratory, Imperial College, London SW7 2AZ, United Kingdom}
    \author{N. Chaturvedi}
	\affiliation{Centre for Inertial Fusion Studies, The Blackett Laboratory, Imperial College, London SW7 2AZ, United Kingdom}
    \author{J. P. Chittenden}
	\affiliation{Centre for Inertial Fusion Studies, The Blackett Laboratory, Imperial College, London SW7 2AZ, United Kingdom}
    \author{B. Duhig}
	\affiliation{Centre for Inertial Fusion Studies, The Blackett Laboratory, Imperial College, London SW7 2AZ, United Kingdom}
    \author{P. W. Moloney}
	\affiliation{Centre for Inertial Fusion Studies, The Blackett Laboratory, Imperial College, London SW7 2AZ, United Kingdom}

    \begin{abstract}
    In inertial confinement fusion, the DT fusion alpha particles carry not only energy but also appreciable momentum that is typically neglected in models of thermonuclear burn. In the central hotspot ignition scheme, the hotspot must self-heat and propagate thermonuclear burn before disassembly. Using radiation–hydrodynamics simulations with a Monte-Carlo alpha particle transport model, we investigate the effect of alpha momentum deposition across sub-ignition to robustly igniting regimes by hydrodynamic scaling of current central hotspot ignition designs from the National Ignition Facility (NIF). We find that the effective alpha particle ram pressure accelerates the shell at burn, reducing hot-spot compression, increasing the rate of disassembly and decreasing yield. This causes a notable ($\sim$ 30\%) reduction in yield at current NIF scale, with a persistent ($\sim$ 10\%) penalty at larger hydrodynamic scales. These results demonstrate that alpha momentum deposition is a significant effect for present ignition-scale implosions, necessitating its inclusion in ignition criteria, burn models, and designs for high-gain inertial confinement fusion.
    \end{abstract}

    \maketitle

    Inertial confinement fusion (ICF) experiments involve the compression of fusion fuel, deuterium and tritium (DT), to high densities and temperatures to achieve thermonuclear burn \cite{Nuckolls1972}. In the central hotspot ignition scheme, a hot, lower density hotspot surrounded by a cold, higher density shell is assembled by implosion. Thermonuclear burn is initiated in the hotspot, and the fusion of DT releases alpha particles which deposit their energy back into the fuel. With sufficient fusion self-heating, the hotspot can `ignite' and propagate a fusion burn wave into the cold shell\cite{Christopherson_2018a,Christopherson2020,Hurricane2014,Hurricane2016,Hurricane2021,Hurricane_PRL2024}. In this manner, ICF implosions can achieve net energy gain from fusion. Ignition and energy gain with the central hotspot ignition scheme has been achieved experimentally at the National Ignition Facility\cite{NIF_PRL2024,Kritcher2022ign,Zylstra2022ign}. For these experiments, fusion particle heating dominates the power balance\cite{Hurricane_PRL2024} and therefore a detailed understanding of the coupling of the fusion products to the fuel is paramount.

    Alpha particle energy deposition is the key physical process in thermonuclear burn and has been thoroughly studied in theory, simulation and current ignition experiments. However, the fast fusion alphas (born with 3.5 MeV of kinetic energy) also carry significant momentum, the effect of which is usually overlooked. For example, for a fusion yield of 1 MJ the alpha particles carry a total momentum of 30 mg km/s. Typical ICF implosion velocities are 300-400 km/s for shell masses of tenths of mg. This clearly indicates that alpha momentum can have a non-negligible impact on burn dynamics.  In this Letter, we show that alpha momentum deposition has a significant deleterious effect on yield for current central hotspot ignition designs. We quantify the effect using detailed alpha transport calculations within radiation hydrodynamics simulations. We show that the largest effect on performance occurs at current NIF experimental scale and the effect is reduced, but persistent, for hydrodynamically scaled implosions.

    The introduction of alpha momentum deposition ($S_{\rho v}$) acts as a source within the momentum equation of hydrodynamics:
    \begin{equation}
        \frac{\partial \rho v}{\partial t} + \nabla \left( P + \rho v^2 \right) = S_{\rho v} \ .
    \end{equation}
    During the stagnation phase of ICF, the flow is subsonic\cite{betti2002} and therefore alpha momentum deposition causes the hotspot and shell to depart from isobaricity. The effective ram pressure of the alphas acts to increase the shell thermal pressure relative to the hotspot pressure. Approximating the alpha momentum deposition term, one can find the shell-hotspot pressure ratio by integrating over the hotspot-shell boundary region\cite{daughton2023} ($\Omega_b$):
    \begin{align}
        \int_{\Omega_b} \nabla P d^3r &\approx 4\pi R_{HS}^2(P_{shell}-P_{HS}) \ , \\
        \int_{\Omega_b} S_{\rho v} d^3r &\approx f_{\rho v} m_\alpha v_\alpha dY/dt \ , \\
        \frac{P_{shell}}{P_{HS}} - 1 &\approx f_{\rho v} \frac{m_\alpha v_\alpha dY/dt}{ 4 \pi P_{HS}R_{HS}^2} \equiv f_{\rho v} \beta \ , \label{eqn:betadef}
    \end{align}
    where $P$ denotes pressure, $R$ radius, $dY/dt$ the total rate of alpha production, and $m_\alpha$ and $v_\alpha$ the mass and velocity of 3.5 MeV DT fusion alpha particles, and we define $\beta$ as the alpha momentum-to-thermal force ratio and $f_{\rho v}$ to be the fraction of total alpha momentum coupled to radial fluid momentum. One can estimate the alpha momentum coupling efficiency, $f_{\rho v}$, from a uniformly-emitting, transparent, spherical hotspot model, for which one can compute $f_{\rho v}$ from the average radial component of alpha velocity at the hotspot edge ($\langle \vec{v}_\alpha \cdot \hat{r} \rangle / v_\alpha$) which is equal to 2/3. The alpha-induced increased pressure gradient across the shell will act to accelerate the shell radially outwards. Radial acceleration of the shell will result in reduced hotspot compression and therefore reduced thermal pressure. Due to fusion yield's strong dependence on hotspot pressure, we therefore expect a reduction in yield in the presence of alpha momentum deposition.

    To compensate for lost compression created by the alpha momentum deposition, the implosion will require additional kinetic energy to achieve ignition conditions. This prompts us to revisit the implosion velocity requirement for ignition and include consideration of the effect of alpha particle momentum. Betti \textit{et al.}\cite{betti2002} derive a hotspot ignition requirement on shell kinetic energy using a compressible shell model. This includes the effects of alpha heating but not momentum deposition i.e. the shell is accelerated by thermal hotspot pressure alone. Starting from Betti's ignition criteria, we add the effective loss in shell velocity due to alpha momentum:
    \begin{subequations}
    \begin{align}
        v^{ign} &= v_{\mathrm{no} \ \alpha \ \rho v}^{ign} + \Delta v_{\alpha \ \rho v}^{ign}  \ , \\
        v_{\mathrm{no} \ \alpha \ \rho v}^{ign} &= 310 \ \mathrm{km/s} \ \left[\alpha_{if}^{2.4} \left(\frac{1 \ \mathrm{mg}}{M_{\mathrm{shell}}}\right) \left(\frac{100 \ \mathrm{MBar}}{P_a}\right)^{0.39}\right]^{\frac{1}{7}} \ , \\
        \Delta v_{\alpha \ \rho v}^{ign} &= \frac{f_{\rho v}Y^{ign}_{\alpha}m_\alpha v_\alpha}{M_{\mathrm{shell}}} = \frac{1}{2} \frac{f_{\rho v}v_{imp}^2 m_\alpha v_\alpha}{Q_{DT}\eta_{\mathrm{hydro}}} \ , \\
        &= 33 \ \mathrm{km/s} \ \left(\frac{ f_{\rho v} }{2/3}\right) \left(\frac{v_{imp}}{400 \ \mathrm{km/s}}\right)^2 \left(\frac{0.05}{\eta_{\mathrm{hydro}}}\right) \ , \nonumber
    \end{align}
    \end{subequations}
    Where the ignition yield ($Y^{ign}_{\alpha}$) is defined as in Hermann, Tabak \& Lindl\cite{HTL2001} (fusion energy equal to absorbed capsule energy), $\alpha_{if}$ is the in-flight adiabat\cite{HTL2001,betti2002}, $M_{\mathrm{shell}}$ is the mass of the shell, $P_{a}$ is the peak ablation pressure, $v_{imp}$ is the implosion velocity and $\eta_{\mathrm{hydro}}$ is the hydrodynamic efficiency\cite{lindl1995}.  Therefore, the implosion velocity required for ignition is increased by the presence of alpha momentum, leading to an effective shift of the `ignition cliff', where the yield is maximally sensitive to implosion parameters.
    
    In the following we will quantify the effect of alpha momentum on current ignition designs, as well as hydrodynamically scaled versions. We use as our baseline the NIF shot N210808\cite{NIF_PRL2024}, the first ICF experiment to exceed Lawson's criterion for ignition\cite{Wurzel2022}. The N210808 design has a peak implosion velocity of $\sim$ 400 km/s as well as a $v_{\mathrm{no} \ \alpha \ \rho v}^{ign} \sim$ 400 km/s. It therefore sits on the ignition cliff. The approximate alpha momentum correction above suggests a few 10s km/s increase in the ignition threshold - this will result in a large change in yield for designs on the ignition cliff. Considering hydrodynamic equivalent\cite{Nora2014} implosions, $v_{imp}^{ign} \propto S^{-3/7}$. Therefore, we expect larger scales to robustly ignite even in the presence of alpha momentum. However, the energy gain will be reduced as alpha momentum deposition will speed up the rarefaction of the burning fuel. To investigate the effects of alpha momentum in different regimes, we simulated hydrodynamic scales, $S$, from 0.75 to 2.0 ($S \sim E_{\mathrm{driver}}^{1/3}$ and therefore equivalent to laser driver energies of 0.8 to 15.2 MJ) to span the sub-ignition to robustly igniting regimes of central hotspot ignition, where we take N210808 as scale 1.
    
    Spherical 1D radiation hydrodynamics simulations were performed with the Eulerian radiation magnetohydrodynamics code Chimera, with 250 nm radial resolution during burn. Chimera has a Monte Carlo alpha heating model\cite{Tong_NF2019,SHERLOCK_JCP2008}, details of ICF implosion modelling with Chimera can be found in the literature\cite{Chittenden2016,Walsh2017,McGlinchey2018,Tong_NF2019,Crilly2022,SpK2022,ONeill_POP2025,Crilly_POP2024}. The as-shot capsule dimensions were used in combination with a frequency dependent spectrum radiation drive (calculated from integrated hohlraum simulations with HYDRA\cite{Marinak2024}) and a surrogate Germanium dopant with increased dopant fraction (tuning found $\sim$ 4x required\cite{ONeill_POP2025}) as compared to the experimentally used Tungsten. A re-tuning of the 210808 design was performed at each new scale factor to ensure hydrodynamic equivalent\cite{Nora2014} implosions. This was performed automatically by a Bayesian optimisation framework\cite{Crilly2025}, aiming to preserve the scale = 1 implosion trajectory parameters ($v_{shell}(t)$, scaled burn-off bang time, scaled areal density) across all scales by varying the thicknesses of the doped and outer undoped layers of HDC ablator.

    To quantify the isolated effect of alpha momentum deposition the Monte Carlo alpha heating model\cite{Tong_NF2019} was run in two modes. Firstly, an alpha energy transport only mode in which alpha particles deposit their energy only and Helium ash is locally deposited, giving rise to fuel depletion effects. Secondly, the full alpha transport mode which includes the mass, momentum and energy transport effects. Note that the effect of spawning alpha particles locally reducing the mass, momentum and energy densities is included. More explicitly, the coupling of alpha particle transport to the hydrodynamics involves the following updates:
    \begin{subequations}
    \begin{align}
        \Delta \rho &= m_{\alpha}n_{\mathrm{thermalise}} - (m_{D}+m_{T})n_{\mathrm{spawn}} \ , \\
        \Delta U_i &= \frac{\sum f_i \Delta E}{V_{\mathrm{cell}}} - \langle E \rangle_{DT} n_{\mathrm{spawn}} \ , \\
        \Delta U_e &= \frac{\sum (1-f_i) \Delta E}{V_{\mathrm{cell}}} \ , \\
        \Delta \rho v &= \frac{m_{\alpha} \sum \Delta v}{V_{\mathrm{cell}}} - (m_{D}+m_{T}) v n_{\mathrm{spawn}} \ , 
    \end{align}
    \end{subequations}
    where $n_i$ is the number density of population $i$, $\Delta E$ and $\Delta v$ are the changes in particle energy and velocity from particle-fluid interactions, $f_i$ is the fraction of energy deposited going to the ions, $\langle E \rangle_{DT}$ is the average energy of reacting DT pairs\cite{Clayton1983}, and summations run over all computational particles in the cell (of volume $V_{\mathrm{cell}}$). The particle energy and velocity changes are computed using both deterministic friction and stochastic diffusion terms, as in Sherlock\cite{SHERLOCK_JCP2008,Tong_NF2019}. The stopping power model is Zimmerman's implementation of Maynard-Deutsch\cite{zimmerman1990recent} and the thermalisation energy cut off is 10 keV. The alpha energy only transport model has $\Delta \rho v = 0$ and $n_{\mathrm{thermalise}} = n_{\mathrm{spawn}}$ for local Helium ash generation. The full transport model includes all terms as above. 

\begin{figure}[htp]
    \centering
    \includegraphics*[width=0.99\columnwidth]{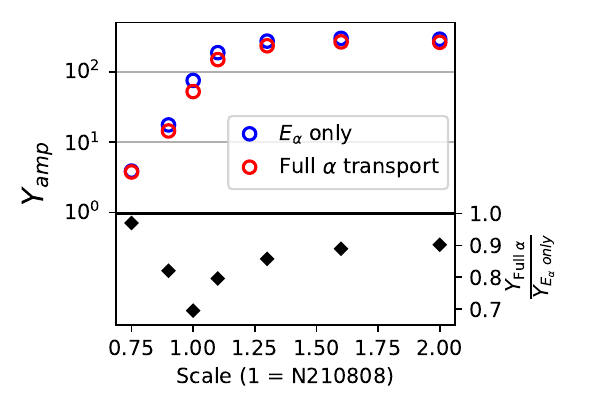}
    \caption{(Top) The yield amplification (ratio of simulated yields with and without alpha heating) from different hydrodynamic scales. Scale = 1 represents current designs at the NIF (in this work, N210808). Shown in blue and red symbols are calculations . (Bottom) The ratio of yields from full alpha particle transport to alpha energy transport only.}
    \label{fig:Yampvsscale}
    \end{figure}

    The scaled radiation hydrodynamics simulations showed that alpha momentum deposition has a significant effect ($\gtrsim$ 10\%) on fusion yield for all ignited implosions ($Y_{amp} > 10$), as shown in \cref{fig:Yampvsscale}. The largest effect was seen at the current NIF scale, where the yield was reduced by $\sim$30\% when the full alpha particle transport model was used. The fusion yields for alpha energy only and full alpha transport models were 4.2 and 2.9 MJ respectively, compared to the 1.37 MJ achieved in the N210808 experiment\cite{Kritcher2022ign,Zylstra2022ign}. The scaled designs plateau in yield amplification at $\sim$300 and $\sim$260 for alpha energy only and full alpha transport models respectively. The effect of alpha momentum shows a similar plateau at a $\sim$10\% reduction in yield. 

    \begin{figure*}[htp]
    \centering
    \includegraphics*[width=0.95\textwidth]{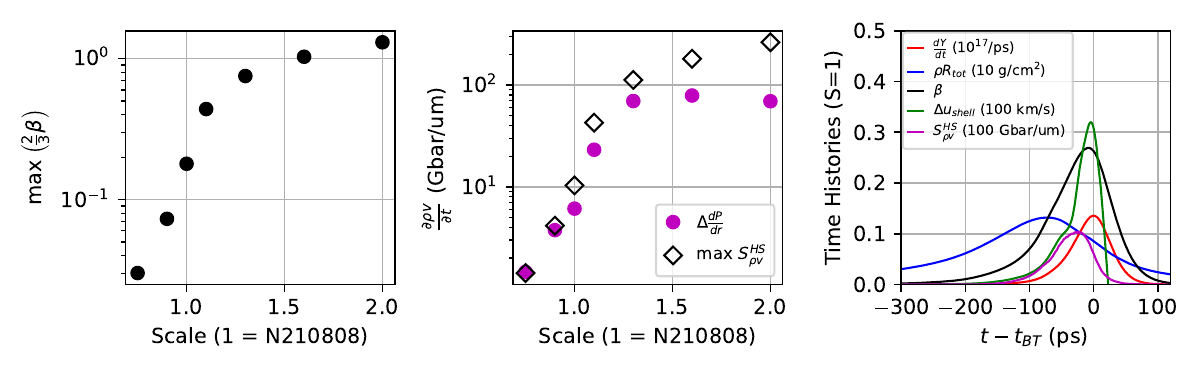}
    \caption{(Left) The maximal value of the alpha momentum-to-thermal force ratio, $\beta$, as a function of hydrodynamic scale. (Middle) The alpha momentum driven pressure gradients and peak alpha momentum deposition rate as a function of scale. The alpha momentum driven pressure gradients, $\Delta dP/dr$, are found by taking the difference between simulations with and without alpha momentum transport. The pressure gradients and alpha momentum deposition rates are calculated at the hotspot radius (defined as the radius enclosing 98\% of neutron production). (Right) Time series from the scale = 1 calculation with the full alpha transport model, where $t_{BT}$ is the nuclear bang time. The change in shell velocity $\Delta u_{\mathrm{shell}}$ is calculated by taking the difference between fuel areal density averaged velocities from simulations with and without alpha momentum transport. }
    \label{fig:PressureEffect}
    \end{figure*}

    The effect of alpha momentum deposition on the dynamics is quantified by the dimensionless force ratio, $\beta$, as defined in \cref{eqn:betadef}. This ratio quantifies the relative importance of alpha momentum generated pressure vs thermal pressure on the acceleration of the shell. As shown in \cref{fig:PressureEffect}, $\beta$ is of order 0.1-1.0 for scales considered in this study. We calculate the alpha momentum coupling efficiency from simulation and we find that $f_{\rho v}$ varies between 0.60 and 0.66, thus $f_{\rho v}$ can be well approximated by the transparent hotspot value of 2/3. The magnitude of the alpha momentum-to-thermal force ratio and momentum coupling efficiency show that alpha momentum will have a notable effect on stagnation and disassembly of the fuel. However, simulations shows that a larger $\beta$ value does not equate to a larger reduction in yield. As the hotspot is igniting, hotspot confinement is crucial to allowing self-heating. Premature hotspot decompression can quench ignition\cite{Hurricane2016}. However, once ignited, burn occurs until the fuel rarefies. This rarefaction occurs at approximately the sound speed\cite{Atzeni2004,Fraley1974} but will be accelerated by the alpha momentum. Fraley \textit{et al.}\cite{Fraley1974} investigated burn propagation in DT microspheres and found that burn fraction tended towards the following function of initial areal density:
    \begin{equation}
        \Phi = \frac{\rho R}{\rho R + H_B} \ , 
    \end{equation}
    where $\Phi$ is the DT fuel burn-up fraction and $H_B$ is the burn parameter, which Fraley \textit{et al.} found to be 6.3 g/cm$^2$ for temperatures between 20 and 70 keV\cite{Fraley1974}. The Chimera simulations for hydrodynamic scales between 1.3 and 2.0 achieved this required temperature range. For these simulation results, we find burn fractions ranging between 27\% - 38\% and 23\% - 34\% for the energy only and full transport models respectively. With peak burn-off $\rho$R as an estimate of initial areal density of an equivalent DT microsphere, we find these burn fractions are consistent with $H_B$ of between 5.5-6.3 g/cm$^2$ and 6.5-7.5 g/cm$^2$ for the energy only and full transport models respectively. This demonstrates the additional difficulty in achieving high burn-up fractions in the presence of alpha momentum driven rarefaction. 

    Alpha momentum deposition will also drive pressure gradients where it is peaked. This is typically at the hotspot edge due to a combination of a geometric effect (the alpha momentum flux is zero by symmetry at the origin) and the rapid decrease in stopping distance at the hotspot edge. We therefore expect to see an increase in radial pressure gradient in the presence of alpha momentum. By finding the time of peak momentum deposition at the hotspot edge, we aim to quantify this effect by finding the difference in the pressure gradient between our two alpha transport model scenarios. As shown in the middle plot of \cref{fig:PressureEffect}, this pressure gradient effect is reflected in the integrated simulations, with increasingly larger positive pressure gradients driven at the peak of momentum deposition. At the larger scales, the change in pressure gradient diverges from the alpha momentum driving force term. However, for these large scales the increased alpha heating in the alpha energy only simulations drives a positive pressure gradient at the hotspot edge. This is because the hotspot becomes transparent and energy deposition becomes peaked at the hotspot edge, resulting in an energy deposition driven pressure gradient.

    For MJ level fusion yields, a simple momentum transfer estimate suggests fusion alpha particles will add 10s km/s of radial velocity to the disassembling fuel. As shown in the right hand plot of \cref{fig:PressureEffect}, the shell velocity is increased by $\sim$30 km/s at the time of peak alpha momentum deposition for scale 1, relative to the alpha energy transport only simulation. This is also the time of peak $\beta$, and therefore the time at which we expect the largest effect from alpha momentum on the shell dynamics.

    \begin{figure}[htp]
    \centering
    \includegraphics*[width=0.99\columnwidth]{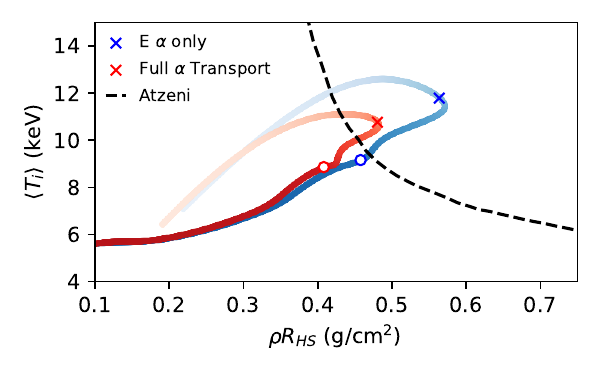}
    \caption{Trajectories in $\rho R_{HS}-\langle T_i \rangle$ space from scale = 1 calculations, full alpha transport in red and energy only in blue. The opacity of the points are coloured by the fraction of burn. Conditions at the time of peak neutron production are shown with crosses and at 10\% of burn with empty circles. Also shown, as a black dashed line, is the ignition boundary as given in Atzeni and Meyer-Ter-Vehn\cite{Atzeni2004}}
    \label{fig:Lawson}
    \end{figure}

    During the onset of ignition at scale 1, the reduced compression during burn caused by the effective alpha particle ram pressure shortens the time for which the hotspot is self-heating. This can be seen by comparing trajectories in $\rho R_{HS}-\langle T_i \rangle$ space\cite{lindl1995,Atzeni2004} with and without alpha momentum deposition effects, as shown in \cref{fig:Lawson}. The reduction in shell velocity, leads to a reduction in hotspot pressure and areal density. The $\rho R_{HS}-\langle T_i \rangle$ trajectory is therefore shifted to lower areal density, entering the ignition region later and therefore with less confinement. At the time when 10 \% of the yield has been produced, alpha ram pressure has reduced the hotspot areal density by 50 mg/cm$^2$. Without this effect, the hotspot is on the point of ignition, by Atzeni's metric\cite{Atzeni2004}. As burn continues, the trajectories diverge considerably, seeded by the $\sim$ 30 km/s reduction in shell velocity. Key measurable quantities reflect the changes in these burn trajectories. Comparing energy only and full alpha transport and experiment\cite{Kritcher2022ign,Zylstra2022ign} respectively, the reduced compression leads to reduced total (0.60 / 0.57 / 0.54 $\pm$ 0.05 g/cm$^2$) and hotspot (0.48 / 0.41 / 0.44 $\pm$ 0.05 g/cm$^2$) fuel areal densities as well as a reduction in the apparent DT (13.9 / 12.1 / 10.9 $\pm$ 0.4 keV) and DD (13.3 / 11.6 / 8.94 $\pm$ 0.4 keV) ion temperatures.

    \begin{figure}[t]
    \centering
    \includegraphics*[width=0.99\columnwidth]{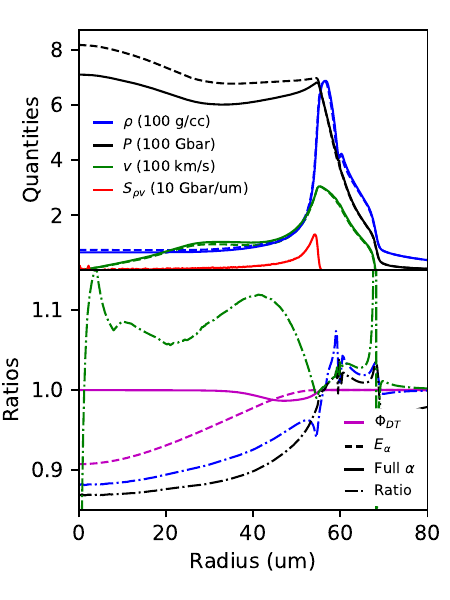}
    \caption{Radial profiles of mass density ($\rho$), pressure ($P$), velocity ($v$), momentum deposition rate ($S_{\rho v}$), local unburnt DT fraction ($\Phi_{DT}$) for the energy only and full alpha transport models, at 10\% of burn. Shown in the dot-dashed lines are ratios of the hydrodynamic quantities where the ratio is (full alpha transport result)/(energy only transport result).}
    \label{fig:Profiles}
    \end{figure}

    Inspecting the hydrodynamic profiles at the 10 \% of burn time for scale 1 shows the spatial dependence of the alpha transport effects, and the same trends discussed below continue throughout the burn. The most notable difference is in the pressure profile. The effect of alpha momentum deposition is a reduction in hotspot pressure and increase in shell pressure. In particular, a positive pressure gradient is established at the hotspot edge, also shown in \cref{fig:PressureEffect}. The alpha momentum deposition is also strongly peaked at the hotspot edge, as discussed above. The positive fluid velocity (radial expansion) is increased across hotspot and shell, apart from at the hotspot edge. At the hotspot edge, the positive pressure gradient acts to decelerate the flow, in line with the expected change to the ablation process\footnote{Extending a steady state ablation analysis\cite{betti2002} to include alpha momentum deposition shows that the induced positive radial pressure gradient also drives faster mass ablation. However, the magnitude of effect is of order $(\rho_{HS}/\rho_{shell})\beta$ and given that the hotspot-shell density contrast is typically large, this effect is second order to shell deceleration.}. Both the reduced hotspot compression and the alpha mass transport effect leads to reduced hotspot density and increased shell density. At this time the effect on temperature is at the few percent level, therefore the majority of the pressure reduction is reflected in the mass density reduction. For the centre of the hotspot, a significant fraction of the density reduction is due to mass transport i.e. non local Helium ash deposition. This is shown as the centre of the hotspot is only 90\% unburnt DT fuel for local mass deposition. However, across the whole hotspot the unburnt DT mass density is reduced in the full transport model, showing that the loss of hotspot areal density is driven predominantly by alpha momentum driven decompression.

    In conclusion, we show that alpha momentum deposition has a significant effect on burn dynamics in the central hotspot ignition scheme. In particular, we show the largest sensitivity ($\sim$ 30\% reduction in yield) on the ignition cliff, where current ICF experiments at the NIF are operating. The alpha momentum acts as effective additional ram pressure which radially accelerates the shell. The increased acceleration results in reduced compression during burn and therefore a loss in fusion yield. This effect is not usually accounted for in theoretical models of ignition and burn\cite{Christopherson_2018a,Christopherson_2018b,daughton2023,Hurricane_PRL2024}. Indeed, a limited number of studies\cite{Galbraith_1991,Kuroki_NF2000,johzaki_2001} carried out, before the National Ignition Facility was built, for early central hotspot ignition designs appear to have been neglected by the community. Without modelling of the deleterious effect of alpha momentum, reduced performance and compression in experiment would likely be attributed to other poorly constrained degradation mechanisms such as fuel preheating, shock mis-timing and fuel-ablator mix\cite{Gaffney2019}. Additionally, momentum deposition (and mass transport) are more difficult to include in alpha energy diffusion models\cite{Atzeni1981,Corman1975} (which are commonplace in ICF simulation codes\cite{Ramis2016,Zimmerman1977,Bellenbaum2024,Woo2018}) compared to particle-based methods\cite{Tong_NF2019,Marinak2024}. In this work, we consider the 1D effect of alpha momentum for central hotspot ignition, however this prompts further investigation into effects on hydrodyanmic instability growth in burn\cite{johzaki_2001}, the creation of alpha induced electrical currents\cite{Appelbe_POP2019, Appelbe_POP2021} and other ignition schemes\cite{Tabak1994,Palaniyappan2025,Cuneo2012}. Looking forward to high gain ICF, the dominant role of fusion products in both energetics and dynamics demands models which include the full complexity of thermonuclear burn.

    \section*{Acknowledgments}
    This research received support through Schmidt Sciences, LLC and the Imperial College Research Fellowship program. This work made use of the Imperial College London RCS High Performance Computing systems.
    
    \section*{References}
    \bibliography{MuCFRefs}

@article{Corman1975,
  title={Multi-group diffusion of energetic charged particles},
  author={Corman, EG and Loewe, WE and Cooper, GE and Winslow, AM},
  journal={Nuclear Fusion},
  volume={15},
  number={3},
  pages={377},
  year={1975},
  publisher={IOP Publishing}
}

@techreport{Zimmerman1977,
  title={LASNEX code for inertial confinement fusion},
  author={Zimmerman, G and Kershaw, D and Bailey, D and Harte, J},
  year={1977},
  institution={California Univ., Livermore (USA). Lawrence Livermore Lab.}
}

@article{Bellenbaum2024,
  title={FLAIM: A reduced volume ignition model for the compression and thermonuclear burn of spherical fuel capsules},
  author={Bellenbaum, Hannah and Read, Martin and Niasse, Nicolas and Barrett, Sean and Hawker, Nicholas and Joiner, Nathan and Chapman, David and others},
  journal={High Energy Density Physics},
  volume={53},
  pages={101159},
  year={2024},
  publisher={Elsevier}
}

@article{Ramis2016,
  title={MULTI-IFE—A one-dimensional computer code for Inertial Fusion Energy (IFE) target simulations},
  author={Ramis, Rafael and Meyer-ter-Vehn, J{\"u}rgen},
  journal={Computer Physics Communications},
  volume={203},
  pages={226--237},
  year={2016},
  publisher={Elsevier}
}

@article{Atzeni1981,
  title={A Diffusive model for $\alpha$-particle energy transport in a laser plasma},
  author={Atzeni, Stefano and Caruso, A},
  journal={Il Nuovo Cimento B (1971-1996)},
  volume={64},
  number={2},
  pages={383--395},
  year={1981},
  publisher={Springer}
}

@article{Cuneo2012,
  title={Magnetically driven implosions for inertial confinement fusion at Sandia National Laboratories},
  author={Cuneo, ME and Herrmann, MC and Sinars, DB and Slutz, SA and Stygar, WA and Vesey, RA and Sefkow, AB and Rochau, GA and Chandler, GA and Bailey, JE and others},
  journal={IEEE Transactions on Plasma Science},
  volume={40},
  number={12},
  pages={3222--3245},
  year={2012},
  publisher={IEEE}
}

@article{Palaniyappan2025,
  title={First indirectly driven liquid-DT filled double shell implosions at the National Ignition Facility},
  author={Palaniyappan, S and Loomis, EN and Negussie, SD and Sauppe, JP and Scott, RL and Robey, HF and Christiansen, NS and Donovan, PM and Wong, CS and Kot, L and others},
  journal={Physics of Plasmas},
  volume={32},
  number={10},
  year={2025},
  publisher={AIP Publishing}
}

@article{Tabak1994,
  title={Ignition and high gain with ultrapowerful lasers},
  author={Tabak, Max and Hammer, James and Glinsky, Michael E and Kruer, William L and Wilks, Scott C and Woodworth, John and Campbell, E Michael and Perry, Michael D and Mason, Rodney J},
  journal={Physics of Plasmas},
  volume={1},
  number={5},
  pages={1626--1634},
  year={1994},
  publisher={American Institute of Physics}
}

@article{Crilly2025,
  title={Automated simulation-based design via multi-fidelity active learning and optimisation for laser direct drive implosions},
  author={Crilly, AJ and Moloney, PW and Shi, D and Ferdinandi, EA},
  journal={arXiv preprint arXiv:2508.20878},
  year={2025}
}

@article{Wurzel2022,
  title={Progress toward fusion energy breakeven and gain as measured against the Lawson criterion},
  author={Wurzel, Samuel E and Hsu, Scott C},
  journal={Physics of Plasmas},
  volume={29},
  number={6},
  year={2022},
  publisher={AIP Publishing}
}

@article{Nuckolls1972,
  title={Laser compression of matter to super-high densities: Thermonuclear (CTR) applications},
  author={Nuckolls, John and Wood, Lowell and Thiessen, Albert and Zimmerman, George},
  journal={Nature},
  volume={239},
  number={5368},
  pages={139--142},
  year={1972},
  publisher={Nature Publishing Group UK London}
}

@article{Fraley1974,
author = {Fraley,G. S.  and Linnebur,E. J.  and Mason,R. J.  and Morse,R. L. },
title = {Thermonuclear burn characteristics of compressed deuterium‐tritium microspheres},
journal = {The Physics of Fluids},
volume = {17},
number = {2},
pages = {474-489},
year = {1974},
doi = {10.1063/1.1694739},

URL = { 
        https://aip.scitation.org/doi/abs/10.1063/1.1694739
    
},
eprint = { 
        https://aip.scitation.org/doi/pdf/10.1063/1.1694739
    
}

}

@article{zimmerman1990recent,
  title={Recent developments in Monte Carlo techniques},
  author={Zimmerman, GB},
  year={1990},
  publisher={Lawrence Livermore National Laboratory (LLNL), Livermore, CA (United States)}
}

@article{Nora2014,
author = {Nora,R.  and Betti,R.  and Anderson,K. S.  and Shvydky,A.  and Bose,A.  and Woo,K. M.  and Christopherson,A. R.  and Marozas,J. A.  and Collins,T. J. B.  and Radha,P. B.  and Hu,S. X.  and Epstein,R.  and Marshall,F. J.  and McCrory,R. L.  and Sangster,T. C.  and Meyerhofer,D. D. },
title = {Theory of hydro-equivalent ignition for inertial fusion and its applications to OMEGA and the National Ignition Facility},
journal = {Physics of Plasmas},
volume = {21},
number = {5},
pages = {056316},
year = {2014},
doi = {10.1063/1.4875331}
}

@article{betti2002,
  title={Deceleration phase of inertial confinement fusion implosions},
  author={Betti, R and Anderson, K and Goncharov, VN and McCrory, RL and Meyerhofer, DD and Skupsky, S and Town, RPJ},
  journal={Physics of Plasmas},
  volume={9},
  number={5},
  pages={2277--2286},
  year={2002},
  publisher={American Institute of Physics}
}

@article{daughton2023,
  title={Influence of mass ablation on ignition and burn propagation in layered fusion capsules},
  author={Daughton, W and Albright, Brian James and Finnegan, Sean Michael and Haines, Brian M and Kline, John L and Sauppe, Joshua Paul and Smidt, Joseph Michael},
  journal={Physics of Plasmas},
  volume={30},
  number={1},
  year={2023},
  publisher={AIP Publishing}
}

@article{HTL2001,
  title={A generalized scaling law for the ignition energy of inertial confinement fusion capsules},
  author={Herrmann, MC and Tabak, M and Lindl, JD},
  journal={Nuclear Fusion},
  volume={41},
  number={1},
  pages={99},
  year={2001},
  publisher={IOP Publishing}
}

@article{Marinak2024,
    author = {Marinak, M. M. and Zimmerman, G. B. and Chapman, T. and Kerbel, G. D. and Patel, M. V. and Koning, J. M. and Sepke, S. M. and Chang, B. and Schroeder, C. R. and Harte, J. A. and Bailey, D. S. and Taylor, L. A. and Langer, S. H. and Belyaev, M. A. and Clark, D. S. and Gaffney, J. and Hammel, B. A. and Hinkel, D. E. and Kritcher, A. L. and Milovich, J. L. and Robey, H. F. and Weber, C. R.},
    title = {How numerical simulations helped to achieve breakeven on the NIF},
    journal = {Physics of Plasmas},
    volume = {31},
    number = {7},
    pages = {070501},
    year = {2024},
    month = {07},
    issn = {1070-664X},
    doi = {10.1063/5.0204710},
    url = {https://doi.org/10.1063/5.0204710},
    eprint = {https://pubs.aip.org/aip/pop/article-pdf/doi/10.1063/5.0204710/20171781/070501_1_5.0204710.pdf},
}

@article{Appelbe_POP2019,
    author = {Appelbe, B. and Sherlock, M. and El-Amiri, O. and Walsh, C. and Chittenden, J.},
    title = {Modification of classical electron transport due to collisions between electrons and fast ions},
    journal = {Physics of Plasmas},
    volume = {26},
    number = {10},
    pages = {102704},
    year = {2019},
    month = {10},
    issn = {1070-664X},
    doi = {10.1063/1.5114794},
    url = {https://doi.org/10.1063/1.5114794},
    eprint = {https://pubs.aip.org/aip/pop/article-pdf/doi/10.1063/1.5114794/15611966/102704_1_online.pdf},
}

@article{Appelbe_POP2021,
    author = {Appelbe, B. and Velikovich, A. L. and Sherlock, M. and Walsh, C. and Crilly, A. and O' Neill, S. and Chittenden, J.},
    title = {Magnetic field transport in propagating thermonuclear burn},
    journal = {Physics of Plasmas},
    volume = {28},
    number = {3},
    pages = {032705},
    year = {2021},
    month = {03},
    issn = {1070-664X},
    doi = {10.1063/5.0040161},
    url = {https://doi.org/10.1063/5.0040161},
    eprint = {https://pubs.aip.org/aip/pop/article-pdf/doi/10.1063/5.0040161/13822896/032705_1_online.pdf},
}

@article{ONeill_POP2025,
    author = {O'Neill, S. T. and Appelbe, B. D. and Crilly, A. J. and Walsh, C. A. and Strozzi, D. J. and Moody, J. D. and Chittenden, J. P.},
    title = {Burn propagation in magnetized high-yield inertial fusion},
    journal = {Physics of Plasmas},
    volume = {32},
    number = {2},
    pages = {022703},
    year = {2025},
    month = {02},
    issn = {1070-664X},
    doi = {10.1063/5.0242215},
    url = {https://doi.org/10.1063/5.0242215},
    eprint = {https://pubs.aip.org/aip/pop/article-pdf/doi/10.1063/5.0242215/20385959/022703_1_5.0242215.pdf},
}

@article{Tong_NF2019,
doi = {10.1088/1741-4326/ab22d4},
url = {https://doi.org/10.1088/1741-4326/ab22d4},
year = {2019},
month = {jun},
publisher = {IOP Publishing},
volume = {59},
number = {8},
pages = {086015},
author = {Tong, J.K. and McGlinchey, K. and Appelbe, B.D. and Walsh, C.A. and Crilly, A.J. and Chittenden, J.P.},
title = {Burn regimes in the hydrodynamic scaling of perturbed inertial confinement fusion hotspots},
journal = {Nuclear Fusion}
}

@article{SHERLOCK_JCP2008,
title = {A Monte-Carlo method for coulomb collisions in hybrid plasma models},
journal = {Journal of Computational Physics},
volume = {227},
number = {4},
pages = {2286-2292},
year = {2008},
issn = {0021-9991},
doi = {https://doi.org/10.1016/j.jcp.2007.11.037},
url = {https://www.sciencedirect.com/science/article/pii/S0021999107005359},
author = {M. Sherlock}
}

@article{Crilly_POP2024,
    author = {Crilly, A. J. and Schlossberg, D. J. and Appelbe, B. D. and Moore, A. S. and Jeet, J. and Kerr, S. and Rubery, M. and Lahmann, B. and O'Neill, S. and Forrest, C. J. and Mannion, O. M. and Chittenden, J. P.},
    title = {Measurements of dense fuel hydrodynamics in the NIF burning plasma experiments using backscattered neutron spectroscopy},
    journal = {Physics of Plasmas},
    volume = {31},
    number = {4},
    pages = {042701},
    year = {2024},
    month = {04},
    issn = {1070-664X},
    doi = {10.1063/5.0203096},
    url = {https://doi.org/10.1063/5.0203096},
    eprint = {https://pubs.aip.org/aip/pop/article-pdf/doi/10.1063/5.0203096/19861719/042701_1_5.0203096.pdf},
}

@article{Kuroki_NF2000,
doi = {10.1088/0029-5515/40/3/306},
url = {https://doi.org/10.1088/0029-5515/40/3/306},
year = {2000},
month = {mar},
publisher = {},
volume = {40},
number = {3},
pages = {357},
author = {Y. Kuroki and Y. Nakao and T. Johzaki and T. Miyahara and H. Nakashima and K. Kudo},
title = {Fusion product momentum deposition  
in laser imploded targets},
journal = {Nuclear Fusion}
}

@article{johzaki_2001, title={Two-dimensional analysis of energy and momentum deposition effects of alpha particles in ICF plasmas}, ISSN={1562-4153}, abstractNote={Two types of two-dimensional codes, i.e. a transport code and a diffusion one, have been developed for analysis of alpha-particle effects in ICF plasmas. On the basis of 2-D coupled transport/hydrodynamic simulations, we investigate the energy and momentum deposition effects of alpha-particles on the Rayleigh-Taylor instability in a stagnating DT planar plasma. After the accuracy validation of the diffusion code, the sensitivity of fuel gain to the perturbation amplitude is also discussed by carrying out 2-D coupled diffusion/hydrodynamic simulations for a DT spherical target with mode number of l=2-12. (author)}, 
journal = {IAEA Conference Proceedings Fusion Energy 2000},
number={no. 8/C}, author={Johzaki, T. and Kuroki, Y. and Nakao, Y.}, year={2001}, month={May}, pages={[5 p.]} }

@article{Galbraith_1991,
author = {David L. Galbraith and Terry Kammash},
title = {Fusion Product Momentum Transfer to Inertially Confined Plasma},
journal = {Fusion Technology},
volume = {19},
number = {3P1},
pages = {492--497},
year = {1991},
publisher = {Taylor \& Francis},
doi = {10.13182/FST91-A29389},
URL = { https://doi.org/10.13182/FST91-A29389},
eprint = {https://doi.org/10.13182/FST91-A29389
}
}

@article{Christopherson_2018a,
    author = {Christopherson, A. R. and Betti, R. and Bose, A. and Howard, J. and Woo, K. M. and Campbell, E. M. and Sanz, J. and Spears, B. K.},
    title = {A comprehensive alpha-heating model for inertial confinement fusion},
    journal = {Physics of Plasmas},
    volume = {25},
    number = {1},
    pages = {012703},
    year = {2018},
    month = {01},
    issn = {1070-664X},
    doi = {10.1063/1.4991405},
    url = {https://doi.org/10.1063/1.4991405},
    eprint = {https://pubs.aip.org/aip/pop/article-pdf/doi/10.1063/1.4991405/14744941/012703_1_online.pdf},
}

@article{Christopherson_2018b,
    author = {Christopherson, A. R. and Betti, R. and Howard, J. and Woo, K. M. and Bose, A. and Campbell, E. M. and Gopalaswamy, V.},
    title = {Theory of alpha heating in inertial fusion: Alpha-heating metrics and the onset of the burning-plasma regime},
    journal = {Physics of Plasmas},
    volume = {25},
    number = {7},
    pages = {072704},
    year = {2018},
    month = {07},
    issn = {1070-664X},
    doi = {10.1063/1.5030337},
    url = {https://doi.org/10.1063/1.5030337},
    eprint = {https://pubs.aip.org/aip/pop/article-pdf/doi/10.1063/1.5030337/14765641/072704_1_online.pdf},
}

@article{Hurricane_PRL2024,
  title = {Energy Principles of Scientific Breakeven in an Inertial Fusion Experiment},
  author = {Hurricane, O. A. and Callahan, D. A. and Casey, D. T. and Christopherson, A. R. and Kritcher, A. L. and Landen, O. L. and Maclaren, S. A. and Nora, R. and Patel, P. K. and Ralph, J. and Schlossberg, D. and Springer, P. T. and Young, C. V. and Zylstra, A. B.},
  journal = {Phys. Rev. Lett.},
  volume = {132},
  issue = {6},
  pages = {065103},
  numpages = {7},
  year = {2024},
  month = {Feb},
  publisher = {American Physical Society},
  doi = {10.1103/PhysRevLett.132.065103},
  url = {https://link.aps.org/doi/10.1103/PhysRevLett.132.065103}
}

@article{NIF_PRL2024,
  title = {Achievement of Target Gain Larger than Unity in an Inertial Fusion Experiment},
  author = {The Indirect Drive ICF Collaboration},
  collaboration = {The Indirect Drive ICF Collaboration},
  journal = {Phys. Rev. Lett.},
  volume = {132},
  issue = {6},
  pages = {065102},
  numpages = {16},
  year = {2024},
  month = {Feb},
  publisher = {American Physical Society},
  doi = {10.1103/PhysRevLett.132.065102},
  url = {https://link.aps.org/doi/10.1103/PhysRevLett.132.065102}
}

@article{Hurricane2016,
  title={Inertially confined fusion plasmas dominated by alpha-particle self-heating},
  author={Hurricane, Omar A and Callahan, DA and Casey, DT and Dewald, EL and Dittrich, TR and D{\"o}ppner, T and Haan, S and Hinkel, DE and Berzak Hopkins, LF and Jones, O and others},
  journal={Nature Physics},
  volume={12},
  number={8},
  pages={800--806},
  year={2016},
  publisher={Nature Publishing Group UK London}
}

@article{Christopherson2020,
  title={Theory of ignition and burn propagation in inertial fusion implosions},
  author={Christopherson, AR and Betti, R and Miller, S and Gopalaswamy, V and Mannion, OM and Cao, D},
  journal={Physics of Plasmas},
  volume={27},
  number={5},
  year={2020},
  publisher={AIP Publishing}
}

@article{Kritcher2022ign,
  title = {Design of an inertial fusion experiment exceeding the Lawson criterion for ignition},
  author = {Kritcher, A. L. and Zylstra, A. B. and Callahan, D. A. and Hurricane, O. A. and Weber, C. R. and Clark, D. S. and Young, C. V. and Ralph, J. E. and Casey, D. T. and Pak, A. and Landen, O. L. and Bachmann, B. and Baker, K. L. and Berzak Hopkins, L. and Bhandarkar, S. D. and Biener, J. and Bionta, R. M. and Birge, N. W. and Braun, T. and Briggs, T. M. and Celliers, P. M. and Chen, H. and Choate, C. and Divol, L. and D\"oppner, T. and Fittinghoff, D. and Edwards, M. J. and Gatu Johnson, M. and Gharibyan, N. and Haan, S. and Hahn, K. D. and Hartouni, E. and Hinkel, D. E. and Ho, D. D. and Hohenberger, M. and Holder, J. P. and Huang, H. and Izumi, N. and Jeet, J. and Jones, O. and Kerr, S. M. and Khan, S. F. and Geppert Kleinrath, H. and Geppert Kleinrath, V. and Kong, C. and Lamb, K. M. and Le Pape, S. and Lemos, N. C. and Lindl, J. D. and MacGowan, B. J. and Mackinnon, A. J. and MacPhee, A. G. and Marley, E. V. and Meaney, K. and Millot, M. and Moore, A. S. and Newman, K. and Di Nicola, J.-M. G. and Nikroo, A. and Nora, R. and Patel, P. K. and Rice, N. G. and Rubery, M. S. and Sater, J. and Schlossberg, D. J. and Sepke, S. M. and Sequoia, K. and Shin, S. J. and Stadermann, M. and Stoupin, S. and Strozzi, D. J. and Thomas, C. A. and Tommasini, R. and Trosseille, C. and Tubman, E. R. and Volegov, P. L. and Wild, C. and Woods, D. T. and Yang, S. T.},
  journal = {Phys. Rev. E},
  volume = {106},
  issue = {2},
  pages = {025201},
  numpages = {14},
  year = {2022},
  month = {Aug},
  publisher = {American Physical Society},
  doi = {10.1103/PhysRevE.106.025201},
  url = {https://link.aps.org/doi/10.1103/PhysRevE.106.025201}
}

@article{Zylstra2022ign,
  title = {Experimental achievement and signatures of ignition at the National Ignition Facility},
  author = {Zylstra, A. B. and Kritcher, A. L. and Hurricane, O. A. and Callahan, D. A. and Ralph, J. E. and Casey, D. T. and Pak, A. and Landen, O. L. and Bachmann, B. and Baker, K. L. and Berzak Hopkins, L. and Bhandarkar, S. D. and Biener, J. and Bionta, R. M. and Birge, N. W. and Braun, T. and Briggs, T. M. and Celliers, P. M. and Chen, H. and Choate, C. and Clark, D. S. and Divol, L. and D\"oppner, T. and Fittinghoff, D. and Edwards, M. J. and Gatu Johnson, M. and Gharibyan, N. and Haan, S. and Hahn, K. D. and Hartouni, E. and Hinkel, D. E. and Ho, D. D. and Hohenberger, M. and Holder, J. P. and Huang, H. and Izumi, N. and Jeet, J. and Jones, O. and Kerr, S. M. and Khan, S. F. and Geppert Kleinrath, H. and Geppert Kleinrath, V. and Kong, C. and Lamb, K. M. and Le Pape, S. and Lemos, N. C. and Lindl, J. D. and MacGowan, B. J. and Mackinnon, A. J. and MacPhee, A. G. and Marley, E. V. and Meaney, K. and Millot, M. and Moore, A. S. and Newman, K. and Di Nicola, J.-M. G. and Nikroo, A. and Nora, R. and Patel, P. K. and Rice, N. G. and Rubery, M. S. and Sater, J. and Schlossberg, D. J. and Sepke, S. M. and Sequoia, K. and Shin, S. J. and Stadermann, M. and Stoupin, S. and Strozzi, D. J. and Thomas, C. A. and Tommasini, R. and Trosseille, C. and Tubman, E. R. and Volegov, P. L. and Weber, C. R. and Wild, C. and Woods, D. T. and Yang, S. T. and Young, C. V.},
  journal = {Phys. Rev. E},
  volume = {106},
  issue = {2},
  pages = {025202},
  numpages = {10},
  year = {2022},
  month = {Aug},
  publisher = {American Physical Society},
  doi = {10.1103/PhysRevE.106.025202},
  url = {https://link.aps.org/doi/10.1103/PhysRevE.106.025202}
}

@article{SpK2022,
  title={SpK: A fast atomic and microphysics code for the high-energy-density regime},
  author={Crilly, AJ and Niasse, NPL and Fraser, AR and Chapman, DA and McLean, KM and Rose, SJ and Chittenden, JP},
  journal={arXiv preprint arXiv:2211.16464},
  year={2022}
}

@article{Crilly2022,
  title={Neutron backscatter edges as a diagnostic of burn propagation},
  author={Crilly, AJ and Appelbe, BD and Mannion, OM and Forrest, CJ and Knauer, JP and Schlossberg, DJ and Hartouni, EP and Moore, AS and Chittenden, JP},
  journal={Physics of Plasmas},
  volume={29},
  number={6},
  pages={062707},
  year={2022},
  publisher={AIP Publishing LLC}
}

@article{Hurricane2021,
author = {Hurricane,O. A.  and Maclaren,S. A.  and Rosen,M. D.  and Hammer,J. H.  and Springer,P. T.  and Betti,R. },
title = {A thermodynamic condition for ignition and burn-propagation in cryogenic layer inertially confined fusion implosions},
journal = {Physics of Plasmas},
volume = {28},
number = {2},
pages = {022704},
year = {2021},
doi = {10.1063/5.0035583}

}

@book{Clayton1983,
  title={Principles of stellar evolution and nucleosynthesis},
  author={Clayton, Donald D},
  year={1983},
  publisher={University of Chicago press}
}

@article{Gaffney2019,
  title={Making inertial confinement fusion models more predictive},
  author={Gaffney, Jim A and Brandon, Scott T and Humbird, Kelli D and Kruse, Michael KG and Nora, Ryan C and Peterson, J Luc and Spears, Brian K},
  journal={Physics of Plasmas},
  volume={26},
  number={8},
  pages={082704},
  year={2019},
  publisher={AIP Publishing LLC}
}

@article{Woo2018,
author = {Woo,K. M.  and Betti,R.  and Shvarts,D.  and Mannion,O. M.  and Patel,D.  and Goncharov,V. N.  and Anderson,K. S.  and Radha,P. B.  and Knauer,J. P.  and Bose,A.  and Gopalaswamy,V.  and Christopherson,A. R.  and Campbell,E. M.  and Sanz,J.  and Aluie,H. },
title = {Impact of three-dimensional hot-spot flow asymmetry on ion-temperature measurements in inertial confinement fusion experiments},
journal = {Physics of Plasmas},
volume = {25},
number = {10},
pages = {102710},
year = {2018},
doi = {10.1063/1.5048429},

URL = { 
        https://doi.org/10.1063/1.5048429
    
},
eprint = { 
        https://doi.org/10.1063/1.5048429
    
}

}

@article{McGlinchey2018,
  title={Diagnostic signatures of performance degrading perturbations in inertial confinement fusion implosions},
  author={McGlinchey, K and Appelbe, BD and Crilly, AJ and Tong, JK and Walsh, CA and Chittenden, JP},
  journal={Physics of Plasmas},
  volume={25},
  number={12},
  pages={122705},
  year={2018},
  publisher={AIP Publishing}
}

@article{lindl1995,
  title={Development of the indirect-drive approach to inertial confinement fusion and the target physics basis for ignition and gain},
  author={Lindl, John},
  journal={Physics of plasmas},
  volume={2},
  number={11},
  pages={3933--4024},
  year={1995},
  publisher={AIP}
}

@book{Atzeni2004,
address = {Oxford},
author = {Atzeni, Stefano and Meyer-ter-Vehn, J{\"{u}}rgen},
booktitle = {International Series of Monographs on Physics},
doi = {10.1093/acprof:oso/9780198562641.001.0001},
isbn = {9780198562641},
language = {eng},
pages = {480},
publisher = {Oxford University Press},
title = {{The Physics of Inertial Fusion: BeamPlasma Interaction, Hydrodynamics, Hot Dense Matter}},
url = {https://www.oxfordscholarship.com/10.1093/acprof:oso/9780198562641.001.0001/acprof-9780198562641},
year = {2004}
}

@article{Hurricane2014,
author = {{Hurricane}, O.~A. and {Callahan}, D.~A. and {Casey}, D.~T. and 
	{Celliers}, P.~M. and {Cerjan}, C. and {Dewald}, E.~L. and {Dittrich}, T.~R. and 
	{D{\"o}ppner}, T. and {Hinkel}, D.~E. and {Hopkins}, L.~F.~B. and 
	{Kline}, J.~L. and {Le Pape}, S. and {Ma}, T. and {Macphee}, A.~G. and 
	{Milovich}, J.~L. and {Pak}, A. and {Park}, H.-S. and {Patel}, P.~K. and 
	{Remington}, B.~A. and {Salmonson}, J.~D. and {Springer}, P.~T. and 
	{Tommasini}, R.},
title = {Fuel gain exceeding unity in an inertially confined fusion implosion},
journal = {Nature},
volume = {506},
pages = {343-348},
year = {2014},

}

@article{Chittenden2016,
author = {Chittenden, J. P. and Appelbe, B. D. and Manke, F. and McGlinchey, K. and Niasse, N. P.L.},
doi = {10.1063/1.4949523},
issn = {10897674},
journal = {Physics of Plasmas},
number = {5},
title = {{Signatures of asymmetry in neutron spectra and images predicted by three-dimensional radiation hydrodynamics simulations of indirect drive implosions}},
volume = {23},
year = {2016}
}

@article{Walsh2017,
author = {Walsh, C. A. and Chittenden, J. P. and McGlinchey, K. and Niasse, N. P.L. and Appelbe, B. D.},
doi = {10.1103/PhysRevLett.118.155001},
issn = {10797114},
journal = {Physical Review Letters},
number = {15},
pages = {1--5},
title = {{Self-Generated Magnetic Fields in the Stagnation Phase of Indirect-Drive Implosions on the National Ignition Facility}},
volume = {118},
year = {2017}
}
	
\end{document}